\documentclass[10pt, aps, pra, reprint, superscriptaddress, nobibnotes]{revtex4-2}

\usepackage{amsmath}
\usepackage{physics}

\usepackage{graphicx}
\usepackage[colorlinks,allcolors=blue]{hyperref}

\begin{document}
\title{Gaussian trajectory description of fragmentation in an isolated spinor condensate}
\author{L. Fernandes}
\email{lennart.fernandes@uantwerpen.be}
\affiliation{Theory of Quantum and Complex Systems, Universiteit Antwerpen, B-2000 Antwerpen, Belgium}

\author{M. Wouters}
\affiliation{Theory of Quantum and Complex Systems, Universiteit Antwerpen, B-2000 Antwerpen, Belgium}

\author{J. Tempere}
\affiliation{Theory of Quantum and Complex Systems, Universiteit Antwerpen, B-2000 Antwerpen, Belgium}
\affiliation{Lyman Laboratory of Physics, Harvard University, Cambridge, Massachusetts 02138, USA}
\date{\today}

\begin{abstract}
Spin-1 Bose gases quenched to spin degeneracy exhibit fragmentation: the appearance of a condensate in more than one single-particle state. Due to its highly entangled nature, the dynamics leading to this collective state are beyond the scope of a Gaussian variational approximation of the many-body wave function. Here, we improve the performance of the Gaussian variational Ansatz by considering dissipation into a fictitious environment, effectively suppressing entanglement within individual quantum trajectories at the expense of introducing a classical mixture of states. We find that this quantum trajectory approach captures the dynamical formation of a fragmented condensate, and analyze how much dissipation should be added to the experiment in order to keep a single realization in a non-fragmented state.
\end{abstract}

\maketitle 

\section{Introduction}
Variational methods are a central tool in theoretical physics, providing a way to approximate the state of a many-body system using a number of parameters much smaller than the full dimension of the Hilbert space.
In particular, time-dependent variational principles (TDVP) allow to probe out-of-equilibrium dynamics with drastically reduced computational resources at the expense of restricting the system's evolution to a small, physically relevant region of the Hilbert space \cite{kramerGeometryTimedependentVariational1981,blaizot1986quantum,haegeman2011time}.

A topic of ongoing interest is the evolution of closed quantum systems after a sudden quench in an external parameter governing the dynamics \cite{polkovnikovColloquiumNonequilibriumDynamics2011,bergesPrethermalization2004,eisertQuantumManybodySystems2015}. 
The early-time dynamics following a quench may be described with great accuracy using a TDVP, as one can usually find a class of variational states which approximates at least the initial condition and states in its vicinity \cite{blaizot1986quantum,haegeman2011time}.
In the long-time limit, equilibration can be proven for a small number of specific cases and asserted more generally for non-integrable systems on the basis of the eigenstate thermalization hypothesis (ETH) \cite{deutsch1991quantum,srednicki1994chaos,rigol2008thermalization,dalessio2016quantum,deutsch2018eigenstate}. While the exact equilibrium state may deviate significantly from any variational state, expectation values of local observables are equivalent to those of a Gibbs distribution of states well suited to a variational approximation \cite{jaynes1957informationI,jaynes1957informationII,rigol2007relaxation}.
At intermediate times, however, no such criterion exists to predict the dynamics of a generic many-body system. Moreover, as the dynamics may carry the system through regions of the Hilbert space far away from those well described by the initial variational Ansatz, the applicability of a single variational theory at all times cannot be guaranteed.

An archetypical variational theory in the context of bosonic many-body systems is the Hartree-Fock-Bogoliubov (HFB) approximation, in which quantum fluctuations are included as Gaussian corrections to a symmetry-breaking wave function \cite{blaizot1986quantum,pitaevskiiBoseEinsteinCondensationSuperfluidity2016}. This approach is predicated on the assumption of a single macroscopically populated single-particle state, perturbed only by small excitations. Bose-Einstein condensation (BEC), the physical phenomenon corresponding to this assumption, has most famously been observed in dilute atomic gases cooled near absolute zero \cite{blochQuantumSimulationsUltracold2012,blochManybodyPhysicsUltracold2008, pitaevskiiBoseEinsteinCondensationSuperfluidity2016}.

In spinor gases, several states of a hyperfine manifold are accessible at comparable energies, introducing an internal degree of freedom and associated spin excitations which, unlike spatial degrees of freedom, are not frozen out by confinement \cite{kawaguchiSpinorBoseEinsteinCondensates2012,stamper-kurnSpinorBoseGases2013}. 
When these spin modes are degenerate, \emph{fragmented} condensation may occur where all hyperfine states are macroscopically populated, requiring a multi-condensate description of the many-body state. This is achieved in the HFB approximation by combining coupled Gross-Pitaevskii equations (GPE) for the wave functions to lowest-order fluctuations dynamics through a Gaussian Ansatz for all spin modes.
However, as the fragmented condensate corresponds to a highly entangled many-body state, an approximation by a single Gaussian state invariably includes large fluctuations \cite{serafiniQuantumContinuousVariables2017, weedbrookGaussianQuantumInformation2012}, invalidating the premise of an approximately coherent state.
In the present work we extend the HFB approximation to describe this fragmentation of the condensate, a phenomenon originally beyond the scope of a Gaussian Ansatz. Taking inspiration from the study of open quantum systems, this will be achieved by considering a fictitious environment into which the spinor gas dissipates. The ensuing classical uncertainty leads to a decomposition of the single squeezed Gaussian state into a classical mixture of states with small fluctuations.
Through this procedure, a more faithful representation of the actual state is formed, capturing the dynamical formation of a fragmented condensate.

The paper is structured as follows: We introduce the single-mode spin-1 gas in Sec. \ref{sec:spinorSMA}, followed by a discussion of spin mixing dynamics in Sec. \ref{sec:spinmixing}. In Sec. \ref{sec:GTA} we introduce the Gaussian trajectory approach, employing it in Sec. \ref{sec:fragmentation} to describe the formation of a fragmented condensate. In Sec. \ref{sec:dissipative} we turn to a dissipative spinor gas and estimate the optimal dissipation rate to observe a single non-fragmented state in experiments. 
We summarize our results in Sec. \ref{sec:conclusions} and provide an outlook for future research.

\section{The single-mode spin-1 gas}
\label{sec:spinorSMA}

Owing to their internal degree of freedom, gases of spin-1 bosons display rich spin dynamics, including the formation of magnetic domains \cite{guzmanLongtimescaleDynamicsSpin2011,jimenez-garciaSpontaneousFormationRelaxation2019, huhObservationStronglyFerromagnetic2020}, fragmented condensation \cite{muellerFragmentationBoseEinsteinCondensates2006,hoFragmentedSingleCondensate2000, evrardProductionCharacterizationFragmented2020} and topological defects \cite{saitoKibbleZurekMechanismQuenched2007,williamsonUniversalCoarseningDynamics2016}.
In contrast to the scalar Bose field $\hat{\psi}(\mathbf{r})$, interactions of the three-component field $\hat{\Psi} = (\hat{\psi}_+ \enspace \hat{\psi}_0 \enspace \hat{\psi}_-)^T$ occur through two separate interaction channels, corresponding to the total angular momentum $S$ of the associated scattering process. Spin-independent density interactions are characterized by $S=0$, whereas processes of $S=2$ comprise interactions in which spin is exchanged among the colliding particles. For dilute gases, both are approximated as contact interactions with interaction strength $g_S = 4\pi \hbar^2 a_S / m$,
where $a_S$ is the s-wave scattering length of the corresponding channel.
The Hamiltonian governing the spin-1 gas is given by \cite{kawaguchiSpinorBoseEinsteinCondensates2012,stamper-kurnSpinorBoseGases2013}
\begin{equation}
    \hat{\mathcal{H}} = \int \dd \mathbf{r} \bigg[\hat{\Psi}^\dagger\bigg( \frac{-\hbar^2 \nabla^2}{2M} + qf_z^2\bigg)\hat{\Psi}
    + \frac{c_0}{2}\hat{n}^2 + \frac{c_2}{2}\hat{\mathbf{S}}^2 \bigg],
    \label{eq:spatial_hamiltonian}
\end{equation}
where $\hat{n}=\hat{\Psi}^\dagger \hat{\Psi}$ is the density operator and $\hat{\mathbf{S}} =\hat{\Psi}^\dagger \mathbf{f}\hat{\Psi}$ the total spin operator, with $\mathbf{f}$ the vector of spin-1 matrices. The quadratic Zeeman splitting $q$ induces a shift in the energy levels of the $\hat{\psi}_\pm$ states as a consequence of a magnetic field applied in the $z$-direction.
The spin-independent and spin-dependent interaction coefficients are given by $c_0 = (g_0 + 2g_2)/3$ and $c_2 = (g_2 - g_0)/3$, respectively. Crucially, the sign of $c_2$ determines the ferromagnetic ($c_2<0$) or antiferromagnetic ($c_2>0$) nature of interactions. We limit our discussion to the antiferromagnetic case.

\begin{figure}[tbp]
    \includegraphics[width=\columnwidth]{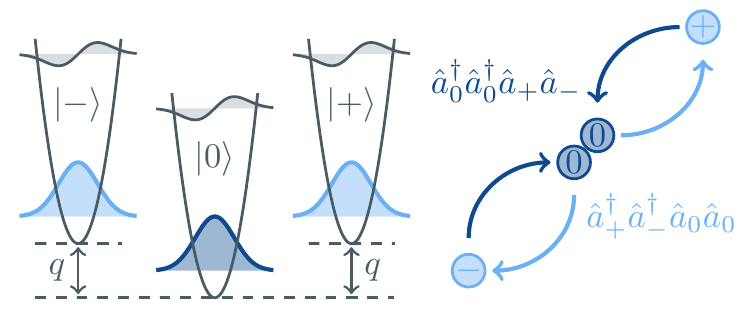}
    \caption{The spin-1 Bose gas reduced to a three-mode spin system in its single-spatial-mode approximation (SMA). The illustrated scattering processes enable the creation and annihilation of spin pairs out of two spin zero particles.}
    \label{fig:illustration_SMA}
\end{figure}

As the scattering lengths $a_0$ and $a_2$ are typically of comparable magnitude \cite{klausenNatureSpinorBoseEinstein2001,stamper-kurnSpinorBoseGases2013}, the interaction strengths satisfy $\abs{c_2} \ll c_0$ and the dynamics are dominated by spin-independent interactions. A perturbational consideration of spin-dependent interactions then leads to the so-called \emph{single-mode approximation} (SMA), in which all hyperfine states are assumed to occupy a shared spatial wave function \cite{lawQuantumSpinsMixing1998, yiSinglemodeApproximationSpinor12002,zhangCoherentSpinMixing2005, sarloSpinFragmentationBose2013}. Within the SMA, the field operators are factorized as $\hat{\psi}_m (\mathbf{r})= \chi(\mathbf{r}) \hat{a}_m$, where $\hat{a}_m(t)$ creates a particle of spin $m$ at time $t$. Both the spatial wave function and the spin populations are normalized through $\int \dd \mathbf{r} \abs{\chi(\mathbf{r})}^2 = 1$ and $\sum_m \expval*{\hat{a}_m^\dagger \hat{a}_m} = N$.
Decoupled from the spatial dynamics, the three mode spin system is then governed by the Hamiltonian
\begin{equation}
    \hat{\mathcal{H}}_S = q\qty(\hat{a}_+^\dagger \hat{a}_+ + \hat{a}_-^\dagger \hat{a}_-)+ \frac{U}{2} \hat{\mathbf{S}}^2,
    \label{eq:spin_hamiltonian}
\end{equation}
where the interaction constant $U = c_2\int \abs{\chi(\mathbf{r})}^4 \dd \mathbf{r}$ depends on the spatial density profile. 
While the SMA is unable to capture the formation of spatial spin structures \cite{jieMeanfieldSpinoscillationDynamics2020}, it remains a valid approximation for small condensates of low magnetization in tight confinement, where domain formation is suppressed \cite{stamper-kurnSpinorBoseGases2013, blackSpinorDynamicsAntiferromagnetic2007,  yiSinglemodeApproximationSpinor12002, puSpinmixingDynamicsSpinor1999,evrardManyBodyOscillationsThermalization2021}.
The single-mode approximation is illustrated in Fig. \ref{fig:illustration_SMA} along with the scattering processes emblematic of the particles' spin nature. Through the last two terms in
\begin{multline}
    \hat{\mathbf{S}}^2 =  \Big(\hat{a}_{+}^\dagger\hat{a}_{+}^\dagger\hat{a}_{+}\hat{a}_{+} + \hat{a}_{-}^\dagger\hat{a}_{-}^\dagger\hat{a}_{-}\hat{a}_{-}
    - 2\hat{a}_{+}^\dagger\hat{a}_{-}^\dagger\hat{a}_{+}\hat{a}_{-} \\
      + 2\hat{a}_{+}^\dagger\hat{a}_{0}^\dagger\hat{a}_{+}\hat{a}_{0} 
      + 2\hat{a}_{-}^\dagger\hat{a}_{0}^\dagger\hat{a}_{-}\hat{a}_{0}\\
      + 2\hat{a}_{0}^\dagger\hat{a}_{0}^\dagger\hat{a}_{+}\hat{a}_{-} + 2\hat{a}_{+}^\dagger\hat{a}_{-}^\dagger\hat{a}_{0}\hat{a}_{0}
     \Big), \label{eq:interactions}
\end{multline}
the creation and annihilation of $\ket{+} \ket{-}$ spin pairs enable a redistribution of particles among spin modes, conserving the total particle number $N$ and angular momentum $\expval*{\hat{S}_z} = \expval*{\hat{a}_+^\dagger \hat{a}_+ - \hat{a}_-^\dagger \hat{a}_-}$ along the $z$-axis. At $\expval*{\hat{S}_z}=0$, these conservation laws restrict the evolution of the system to a subspace of dimension $N/2 +1$, allowing for an exact solution of the many-body Schr\"odinger equation \cite{lawQuantumSpinsMixing1998, miasQuantumNoiseScaling2008}, detailed in Appendix \ref{appendix:hamiltonian}. 
As we will restrict our discussion to $\expval*{\hat{S}_z}=0$, the fraction of atoms in the $\ket{\pm}$ modes remains equal at all times and will be denoted as the (fractional) \emph{pair number} $n_p = \expval*{\hat{a}_\pm^\dagger \hat{a}_\pm}/N$. This quantity takes on values in the range $\left[ 0, 1/2 \right]$ and will be the main observable of interest for the remainder of this paper.

\begin{figure}[tbp]
    \includegraphics[scale=1]{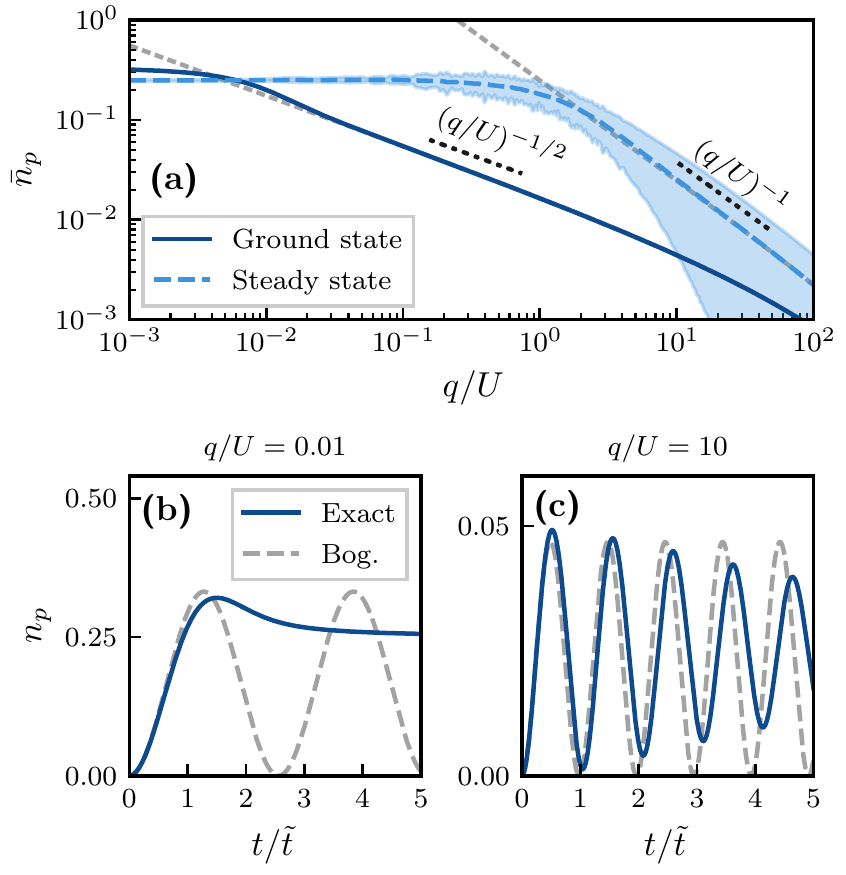}
    \caption{Ground state and dynamical phase diagram for an antiferromagnetic spin-1 gas of $N=400$ and $\expval*{\hat{S}_z}=0$. (a) The mean number of spin pairs in the ground state and the nonequilibrium steady state reached through a quench from $q_i=\infty$ to $q_f=q$, both calculated with the exact many-body Schr\"odinger equation. The shaded region around the latter represents the residual oscillation amplitude. Grey dashed lines indicate the corresponding Bogoliubov predictions. (b-c) Typical dynamics of spin populations for quenches to values of $q/U$ in two distinct regimes.}
    \label{fig:phasediagram}
\end{figure}

Both the ground state and dynamical behavior of the spin-1 gas exhibit interesting properties as a function of the dimensionless parameter $q/U$. Fig. \ref{fig:phasediagram}(a) shows the pair number in the ground state of a gas for different values of $q/U$. 
For $q=0$, the three single-particle orbitals are degenerate and equally populated, $\bar{n}_p=1/3$. As the Zeeman splitting increases, the occupation of the $\ket{\pm}$ modes is suppressed and most particles condense in the $\ket{0}$ mode. 
The scaling at high $q/U$ can be derived in a Bogoliubov approximation of small spin excitations on top of a $\ket{0}$ condensate, as detailed in Appendix \ref{appendix:bogoliubov}.
At lower $q/U$, both the dynamical and ground state predictions of Bogoliubov theory become inaccurate due to a non-negligible occupation of the $\ket{\pm}$ modes.

\section{Spin mixing dynamics}
\label{sec:spinmixing}
A sudden quench in the Zeeman splitting abruptly changes the ground state of the gas, allowing to probe the out-of-equilibrium dynamics as it evolves toward its new steady state. We consider the dynamics of a gas prepared in a coherent state $\ket{0}^{\bigotimes N}$ (i.e. the ground state for $q/U \to \infty$), quenched to a finite value of $q/U$.
Analogously to the ground state, the ensuing spin mixing dynamics can be categorized into two regimes, separated by a continuous crossover.
For $q/U \gg 1$, the dynamics is dominated by the quadratic Zeeman contribution to the Hamiltonian \eqref{eq:spin_hamiltonian}. The time evolution is therefore close to linear and the number of spin pairs exhibits persistent small amplitude coherent oscillations \cite{changCoherentSpinorDynamics2005, wideraCoherentCollisionalSpin2005, gerbierResonantControlSpin2006, huhObservationStronglyFerromagnetic2020, evrardManyBodyOscillationsThermalization2021,evrardCoherentSeedingDynamics2021}, illustrated in Fig. \ref{fig:phasediagram}(c). As shown in the figure, the early time dynamics in this regime are adequately captured by Bogoliubov theory.
On the other hand, for $q/U \ll 1$, interactions dominate the dynamics and the pair number relaxes rapidly to its steady state, shown in Fig. \ref{fig:phasediagram}(b). For negligible $q/U$, the steady state corresponds to a fragmented condensate with $\bar{n}_p=1/4$, reached on a timescale $\tilde{t}=\hbar / (U\sqrt{2N})$ \cite{lawQuantumSpinsMixing1998,evrardManyBodyOscillationsThermalization2021}. Different from the ground state configuration $\bar{n}_p=1/3$, this value corresponds to the equilibrium expectation value in a generalized Gibbs ensemble (GGE) which accounts for the conservation of $\hat{S}_z$ \cite{evrardManyBodyOscillationsThermalization2021}.
The dashed line in Fig. \ref{fig:phasediagram}(a) indicates the mean pair number in the steady state across the $q/U$ parameter range, while the shaded region around it represents the amplitude of persistent oscillations.

\begin{figure}[tbp]
    \includegraphics[scale=1]{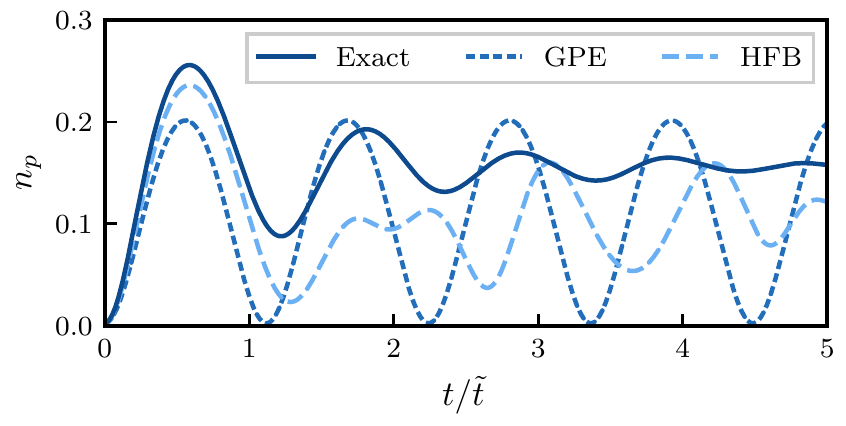}
    \caption{Gaussian HFB dynamics at $q/U=10$ for a gas of $N=400$ atoms and seeded initial condition $N_{\textnormal{seed}}=1$. The exact solution and a mean field prediction are shown for comparison.}
    \label{fig:HFB_dynamics}
\end{figure}

As the Bogoliubov approximation breaks down when multiple spin modes become macroscopically occupied, we employ a more general Gaussian Ansatz in which all three fields $\hat{a}_m$ are described through their first and second moments. This comes down to an expansion 
\begin{equation}
    \hat{a}_m = \phi_m + \hat{\delta}_m,
    \label{eq:expansion}
\end{equation}
decomposing the fields into a condensate mode $\phi_m = \expval*{\hat{a}_m}$ and fluctuations of zero mean, whose Gaussian nature is reflected in the Wick decomposition of higher order moments into products of quadratic correlations $\expval*{\hat{\delta}_m^{(\dagger)} \hat{\delta}_n}$. Coupled Gross-Pitaevskii equations for the fields $\phi_m$ and Heisenberg equations of motion for the quadratic correlations then form a closed system known as the Hartree-Fock-Bogoliubov (HFB) approximation \cite{blaizot1986quantum, colussiCumulantTheoryUnitary2020}, detailed in Appendix \ref{appendix:HFB}.
While in principle suited to describe macroscopic occupation of all modes, the creation of spin pairs out of a coherent $\ket{0}^{\bigotimes N}$ initial state results only in a growth of fluctuations, leaving the $\phi_m$ modes unoccupied. The HFB theory therefore coincides with the Bogoliubov predictions in Fig. \ref{fig:phasediagram}(b-c). This can be partially resolved by considering coherently seeded dynamics \cite{evrardCoherentSeedingDynamics2021}, where a small number of atoms are prepared as $N_{seed}$ spin pairs in the $\phi_\pm$ modes. As shown in Fig. \ref{fig:HFB_dynamics}, this activation of the coupled GPE's causes a deviation from the oscillations predicted by Bogoliubov theory, but still fails to reproduce relaxation to the fragmented state.

The fundamental limitation of the Gaussian theory lies in the growth of fluctuations, which over time exceed the occupation of the mean field modes.
As the expansion \eqref{eq:expansion} is predicated on a limited depletion $\expval*{\hat{\delta}_m^\dagger \hat{\delta}_m} \ll \abs{\phi_m}^2$, a Gaussian state dominated by fluctuations can no longer be expected to form an accurate representation of the actual state of the system.
For comparison, we include in Fig. \ref{fig:HFB_dynamics} also a mean field approximation which considers only the coupled GPE's, thus requiring a seeded initial condition to produce spin mixing dynamics. Mean field theory captures the creation of spin pairs at early times, but, like Bogoliubov theory, is limited to coherent oscillations \cite{zhangCoherentSpinMixing2005, heinzeInfluenceParticleNumber2010}.

\section{A Gaussian trajectory approach}
\label{sec:GTA}

The growth of fluctuations in the Gaussian approximation is a consequence of growing entanglement during the formation of a fragmented condensate. To overcome the entanglement hampering the Gaussian theory, we take inspiration from equilibrium statistical mechanics. There, actual eigenstates may be highly entangled, but the expectation values of local operators are equivalent to those of a mixed state of which the density matrix obeys a Gibbs distribution $\rho \propto e^{-\beta \hat{H}}$, which can be approximated as a mixture of minimally entangled states \cite{stoudenmire2010minimally}.
To find an approximate representation of a single entangled nonequilibrium state as a classical mixture of weakly entangled states, we resort to the quantum trajectory framework \cite{breuerTheoryOpenQuantum2002,daleyQuantumTrajectoriesOpen2014}. This method for the description of open quantum systems samples the time evolution of a density matrix through stochastic realisations of a single quantum state, known as \emph{unravelings} of the underlying master equation \cite{verstraelenGaussianQuantumTrajectories2018}. In the context of open quantum systems, the stochastic nature and the ensuing classical uncertainty derive from the coupling to an environment, which continuously performs measurements on the system. In our case, classical uncertainty is artificially introduced to approximate the entangled, yet pure, many-body state.

\begin{figure}[tbp]
    \includegraphics[width = \columnwidth]{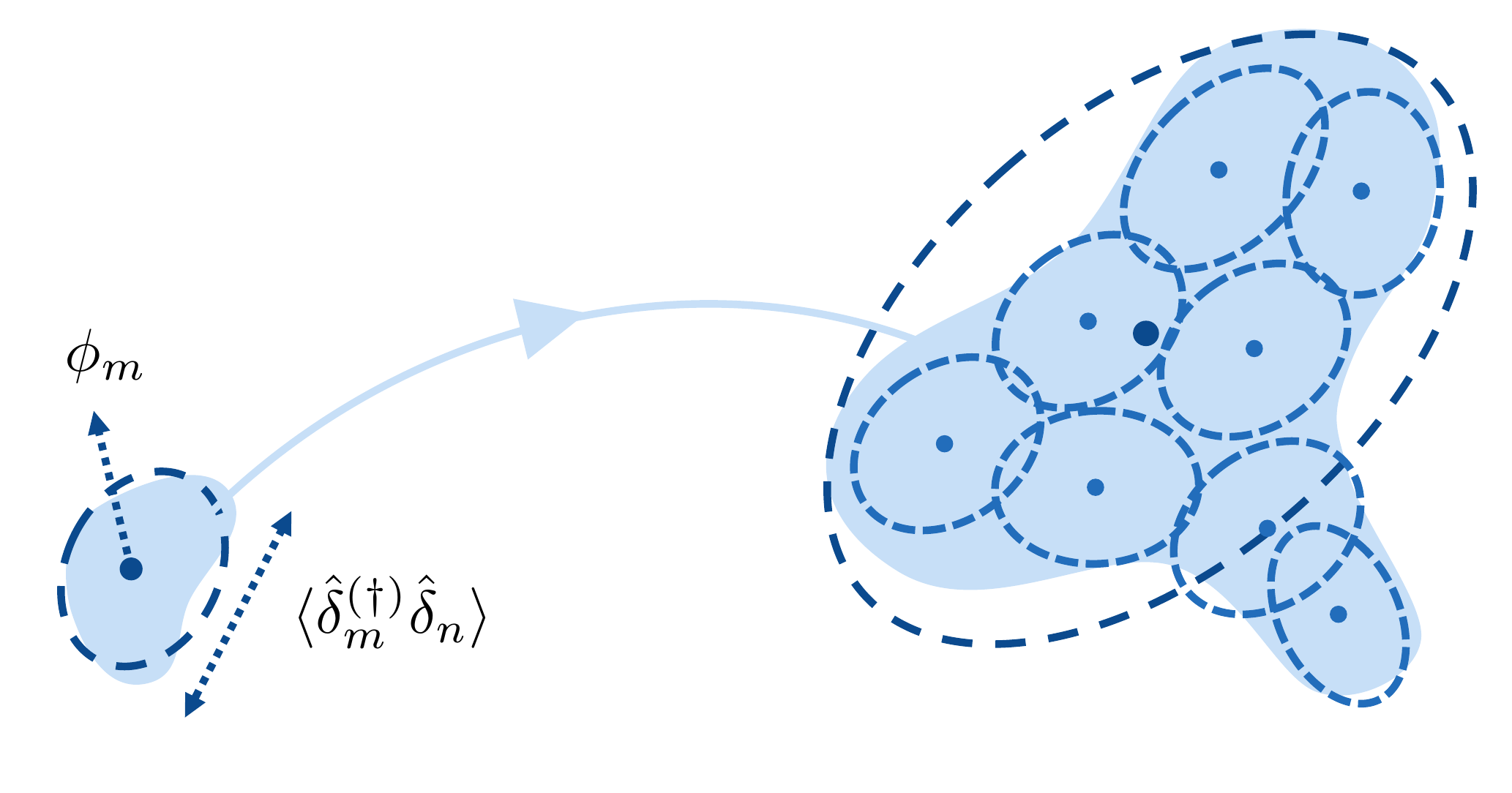}
    \caption{Depiction of the Gaussian trajectory approach to quench dynamics. While a single variational state consisting of a mean field mode $\phi_m$ and Gaussian fluctuations $\hat{\delta}_m$ cannot capture the highly non-Gaussian final state, a classical mixture of variational states with small fluctuations produces a more accurate representation.}
    \label{fig:GTA_sketch}
\end{figure}

Considering single-particle losses ($\hat{\Gamma}_m = \sqrt{\gamma} \hat{a}_m$) under heterodyne unraveling of the Lindblad master equation, the time evolution of an expectation value in a single trajectory is given by \cite{verstraelenGaussianQuantumTrajectories2018}
\begin{multline}
    \dd \expval*{\hat{O}} = \, - i \expval{\commutator{\hat{O}}{\hat{\mathcal{H}}_S}} \dd t \\
    - \frac{\gamma}{2}\sum_m \qty( \expval{\anticommutator{\hat{a}_m^\dagger \hat{a}_m}{\hat{O}}} - 2 \expval{\hat{a}_m^\dagger \hat{O} \hat{a}_m} )\dd t \\
    + \sqrt{\gamma} \sum_m \qty( \expval{\hat{a}_m^\dagger(\hat{O}- \expval*{\hat{O}})}\dd Z_m+ c.c. ).
    \label{eq:heterodyne_lindblad}
\end{multline}
Here, the Hamiltonian evolution is supplemented with a deterministic ($\propto \gamma \dd t$) and stochastic ($\propto \sqrt{\gamma} \dd Z_m$) dissipation term, where the latter contains uncorrelated complex Wiener processes satisfying $\expval*{\dd Z_m^* \dd Z_n}=\delta_{m,n}\dd t$.
Referring to Appendix \ref{appendix:HFB} for the full equations, the effect of dissipation on the time evolution of the Gaussian moments can be summarized as
\begin{align}
    \frac{\dd}{\dd t} \expval*{\abs{\phi_m}^2} &\sim -\gamma \abs{\phi_m}^2 + \gamma \sum_n  \abs*{\expval*{\hat{\delta}_m^{(\dagger)} \hat{\delta}_n}}^2 , \label{eq:heterodyne_phi}\\
    \frac{\dd}{\dd t} \expval*{\hat{\delta}_m^{\dagger} \hat{\delta}_m} &\sim -\gamma \expval*{\hat{\delta}_m^{\dagger} \hat{\delta}_m} - \gamma \sum_n \abs*{\expval*{\hat{\delta}_m^{(\dagger)} \hat{\delta}_n}}^2 ,
    \label{eq:heterodyne_delta}
\end{align}
where the former result requires the application of It\={o}'s lemma. Comparing the last terms in \eqref{eq:heterodyne_phi} and \eqref{eq:heterodyne_delta} shows that the predominant part of the deterministic decrease in fluctuations is on average compensated by a stochastic growth of the condensate modes.
The net result of adding dissipation is thus an effective conversion of quantum into classical fluctuations, hereby converting quantum superpositions into classical uncertainty  while approximately conserving the total amount of fluctuations \cite{woutersQuantumTrajectoriesVariational2020}.
The reduction of fluctuations at the expense of introducing a classical mixture allows one to form a composite Gaussian approximation of a single highly non-Gaussian many-body state. This method was recently implemented to describe chaotic dynamics in a 4-site Bose-Hubbard chain, where a similar growth of entanglement hinders a variational description in terms of a single Gaussian state \cite{woutersQuantumTrajectoriesVariational2020}. As illustrated in Fig. \ref{fig:GTA_sketch}, each state constituting the mixture corresponds to a single quantum trajectory and is captured by the Gaussian HFB theory due to the suppressed amplitude of fluctuations. 

In the case of open systems, the inferred mixed state corresponds to the actual density matrix of the system. For the present purpose of describing a closed system, dissipation into a fictitious environment is an artificial construction to obtain an approximate representation of a single highly entangled state as a classical mixture of weakly entangled states.
Consequently, the rate $\gamma$ is no property of the actual system and cannot be based on physical considerations. We therefore allow the isolated system to evolve undisturbed according to its Hamiltonian \eqref{eq:spin_hamiltonian} until the total number of fluctuations $\Delta = \sum_m \expval*{\hat{\delta}_m^\dagger \hat{\delta}_m}$ reaches an upper critical value $\Delta_{\text{c}}$. The Hamiltonian evolution is then halted and the system at instantaneous time evolved according to the dissipative part of the dynamics (the second and third lines in \eqref{eq:heterodyne_lindblad}) until fluctuations are suppressed below $\Delta_s = \Delta_{\text{c}}/2$. 
The main advantage of this two-step procedure is that the dissipation evolves adaptively during the time evolution of the system, according to the growth of fluctuations at any given time. This allows to retain a Hamiltonian time evolution minimally affected by the artificial dissipation. 
As seen from the first terms in \eqref{eq:heterodyne_phi} and \eqref{eq:heterodyne_delta}, the added dissipation breaks the conservation of particle number and angular momentum $\expval*{\hat{S}_z}$. In the proposed scheme, this deviation is corrected by a rescaling of the amplitudes $\abs{\phi_m}$ to enforce $\abs{\phi_+} = \abs{\phi_-}$ and $\sum_m (\abs{\phi_m}^2 + \expval*{\hat{\delta}_m^\dagger \hat{\delta}_m}) = N$ following each dissipation step, projecting the evolved state back onto the variational manifold before Hamiltonian time evolution resumes.

\section{Dynamics of fragmentation}
\label{sec:fragmentation}

\begin{figure}[tbp]
    \includegraphics[scale=1]{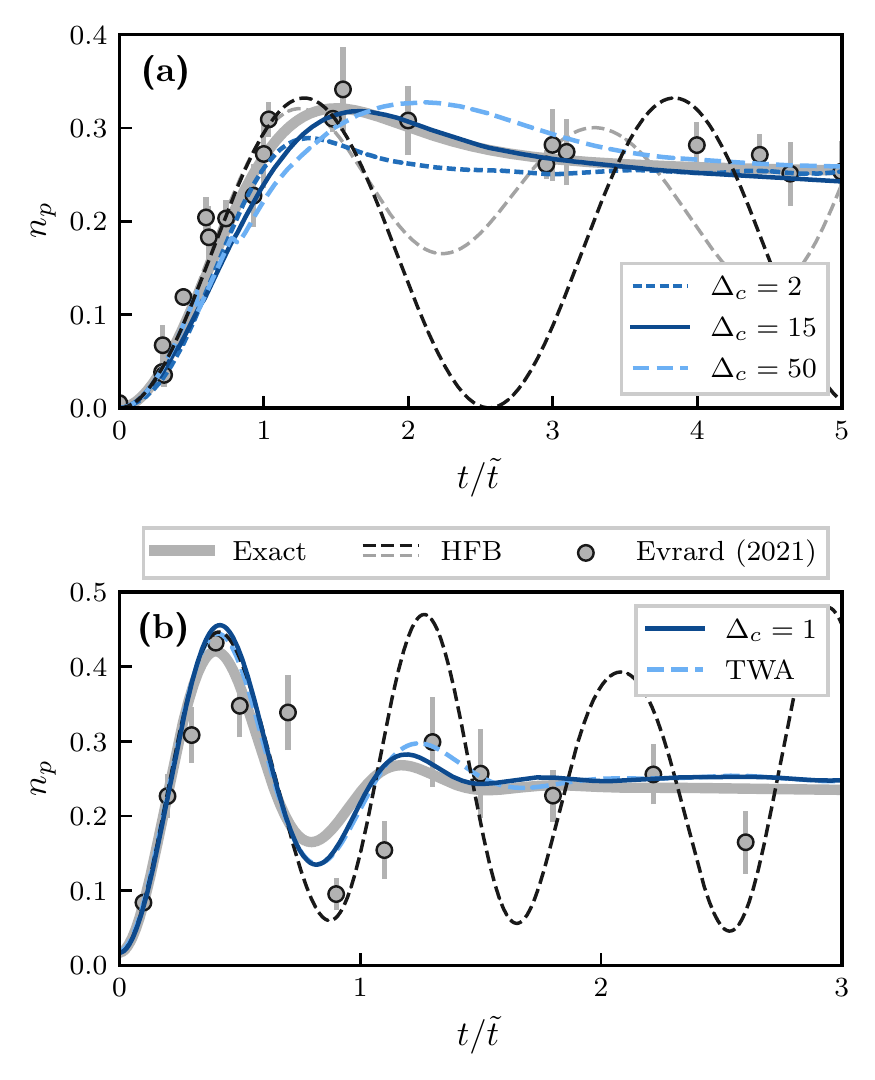}
    \caption{Rapid relaxation of the pair number in the interaction-dominated regime ($q/U = 0$), showing the improvement of the variational description due to the added dissipation. Results are shown for a gas of $N=200$ atoms initiated in the $\ket{0}^{\bigotimes N}$ coherent state with $N_{\textnormal{seed}}=0$ (a) and $N_{\textnormal{seed}}=3.4$ (b). The grey dashed line in (a) represents the HFB result for $N_{\textnormal{seed}}=10^{-2}$, illustrating its sensitivity to initial conditions. Gaussian trajectory and TWA results represent the average over $10^4$ realisations. Experimental data were taken from \cite{evrardManyBodyOscillationsThermalization2021, evrardCoherentSeedingDynamics2021}. }
    \label{fig:Plot_Relaxation}
\end{figure}

Fig. \ref{fig:Plot_Relaxation}(a) shows the evolution of the number of spin pairs after an instantaneous quench of a coherent $\ket{0}^{\bigotimes N}$ state into the interaction-dominated regime $q/U = 0$, where both the exact solution and experimental results \cite{evrardManyBodyOscillationsThermalization2021} reveal a rapid relaxation to the fragmented state. While the HFB approximation (black dashed line) accurately predicts the onset of spin mixing, it fails to reproduce the eventual relaxation, instead predicting perpetual oscillations of the pair number. For comparison, we show in grey the same HFB result for an inifinitesimal seeding $N_{seed}=10^{-2}$, demonstrating the sensitivity of the HFB result to the initial conditions of the gas. 
The Gaussian trajectory method eliminates this instability as the introduction of classical mixedness provides a natural ensemble average.
Both the early growth of spin pairs and the eventual final state appear robust with respect to the chosen value of $\Delta_c$, its effect being most pronounced where the pair number reaches its maximum. In this intermediate regime, we find that a value of $\Delta_c \approx 15$ best replicates the exact result when fluctuations are suppressed to $\Delta_s = \Delta_c /2$ in each dissipation step. 
For this simulation, the exact result is indistinguishable from a truncated Wigner approximation (TWA) \cite{sinatraTruncatedWignerMethod2002,blakieDynamicsStatisticalMechanics2008, polkovnikovPhaseSpaceRepresentation2010}, which approximately includes quantum fluctuations in the mean field equations through a stochastic sampling of initial conditions in accordance with the Wigner distribution of the coherent state.
The same quench scenario is shown in Fig. \ref{fig:Plot_Relaxation}(b) for a coherently seeded initial state, where about $3.4 \%$ of the atoms are prepared as spin pairs, activating the coupled mean field equations in HFB theory. The seeded time evolution is marked by a longer persistence of oscillations, captured by the HFB approximation at early times. At intermediate times, the TWA prediction slightly underestimates the damping rate of oscillations. We find the Gaussian trajectory result optimally replicates the relaxation dynamics for $\Delta_c\leq 1$. Below this value, the result is independent on $\Delta_c$, while a larger $\Delta_c$ causes an overestimation of the oscillation amplitude. In the optimal regime, the Gaussian trajectory approach slightly outperforms the TWA prediction.

Our initial objective motivating the introduction of quantum trajectories was to obtain an approximate representation of a single, highly entangled state as a classical mixture of weakly entangled states. A hallmark of entanglement in a pure many-body state is inseparability, meaning that the reduced state of a constituent subsystem cannot be represented by a pure state \cite{serafiniQuantumContinuousVariables2017}. This is also the case for the spin-1 gas, for which the single-particle reduced density matrix  $\rho^{(n,m)}_{1} = \expval*{\hat{a}_n^\dagger \hat{a}_m}/N$ is given by $\textnormal{diag}(n_p, 1-2n_p, n_p)$ due to the symmetries of the Hamiltonian (see Appendix \ref{appendix:hamiltonian}).
Fragmentation of the condensate at high $n_p$ is reflected precisely in the diagonal nature of the reduced density matrix, which at high $n_p$ describes a macroscopic occupation of the individual single-particle orbitals rather a single condensate in a superposition of all three \cite{evrardProductionCharacterizationFragmented2020}. 

\begin{figure}[tbp]
    \includegraphics[scale=1]{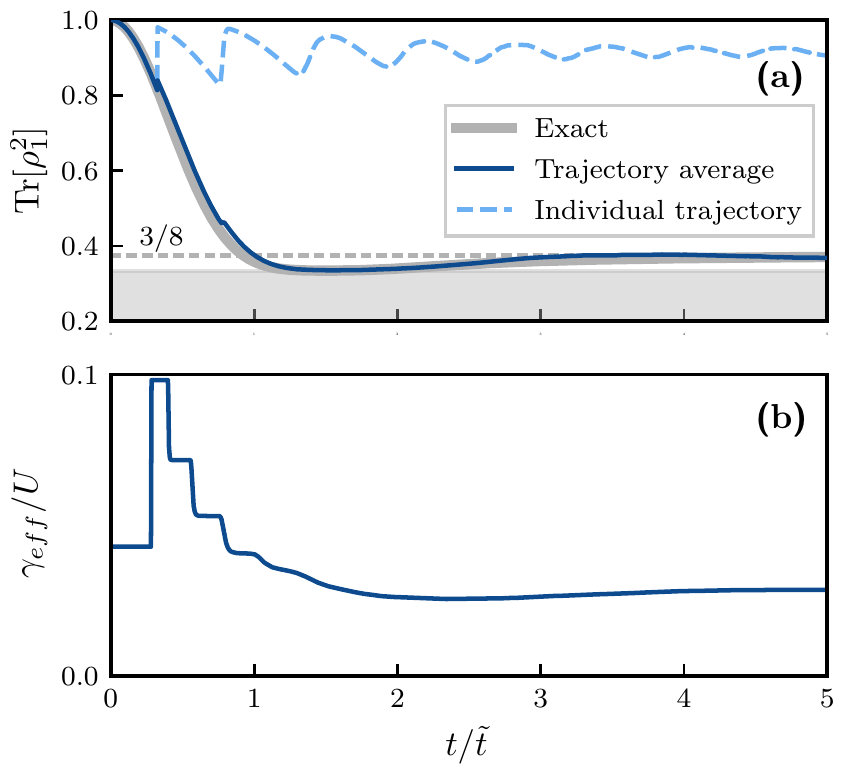}
    \caption{Growth of entanglement during the dynamics of Fig. \ref{fig:Plot_Relaxation}(a). (a) Evolution of the single-particle reduced density matrix's purity. The sudden jumps in the purity of individual trajectories correspond to instantaneous dissipation steps. To emphasize the suppression of fluctuations, the result is shown for $\Delta_s = \Delta_c /10$, leading to an optimal critical value of $\Delta_c = 20$. (b) Effective dissipation rate $\gamma_{eff}$ inferred from the Gaussian trajectory result.}
    \label{fig:Plot_Entanglement}
\end{figure}

The mixedness of $\rho_1$, measured by the purity $\Trace [\rho_1^2]$, serves as a measure of entanglement in the many-body state.
Fig. \ref{fig:Plot_Entanglement}(a) shows the evolution of the purity of $\rho_1$ during a quench into the fragmented state. 
The steady state value $3/8$ reproduced by the trajectory average corresponds to the fragmented condensate with relative populations $(1/4, 1/2, 1/4)$. By contrast, the suppression of Gaussian fluctuations within a single trajectory enhances the purity, evidencing the reduction of entanglement.
Note that the Gaussian trajectory \emph{average} corresponds to an actual mixture of variational states. The mixedness of the reduced density matrix is affected by this classical uncertainty and is therefore no direct measure of entanglement. The agreement between the purity in the trajectory average and exact solution testifies to the conservation of total (classical and quantum) fluctuations by the trajectory approach, as illustrated in Fig. \ref{fig:GTA_sketch}.
For the exact solution as well as within a single variational trajectory, the reduced density matrix's purity does qualify as an indication for entanglement, as both always correspond to a pure many-body state \cite{serafiniQuantumContinuousVariables2017, guhneEntanglementDetection2009}.

The generation of entanglement during the formation of the fragmented condensate is also witnessed in the dissipation rate of the Gaussian trajectory approach. Since the adaptive dissipation procedure described in Sec. \ref{sec:GTA} suppresses fluctuations below an upper critical value $\Delta_c$, the required dissipation strength to do so may vary throughout a single trajectory, as the generation of entanglement does not necessarily occur at a constant rate.
To illustrate this, we show in Fig. \ref{fig:Plot_Entanglement}(b) the average effective dissipation rate throughout the dynamics of Fig. \ref{fig:Plot_Relaxation}(a). The effective dissipation rate $\gamma_{eff}$ is derived by relating the number $N_\gamma$ of instantaneous dissipation steps of numerical size $\gamma dt'$ to the time $\Delta t$ of the free Hamiltonian evolution preceding it:  
\begin{equation}
    \gamma_{eff}=  N_\gamma \gamma dt' / \Delta t.
\end{equation}
At early times, fluctuations grow rapidly and a high dissipation rate is required to suppress the generated entanglement through frequent stochastic jumps. At late times the procedure evolves towards a lower dissipation rate as entanglement saturates in the fragmented state. Due to the variable generation of entanglement, a constant continuous dissipation rate does not capture the fragmentation dynamics as well as the two-step method with explicitly constrained fluctuations.

\begin{figure}[tbp]
    \centering
    \includegraphics{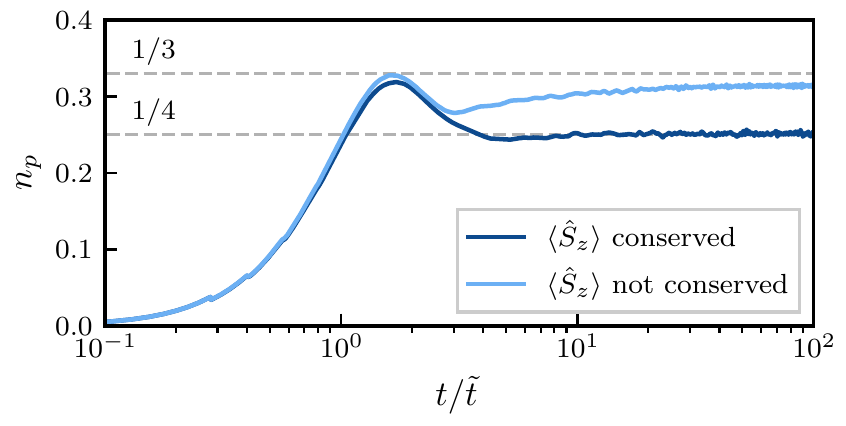}
    \caption{Long time simulation of the relaxation pictured in Fig. \ref{fig:Plot_Relaxation}(a), showing the effect of $\expval*{\hat{S}_z}$ conservation within single trajectories.}
    \label{fig:conservation_law}
\end{figure}

As mentioned in Sec. \ref{sec:spinorSMA}, the steady state pair number $\overline{n}_p=1/4$ in the fragmented condensate is a consequence of $\expval*{\hat{S}_z}$ conservation which inhibits the system from reaching thermal equilibrium \cite{evrardManyBodyOscillationsThermalization2021}. In the trajectory simulation of the isolated gas, this conservation law is fixed by restoring $\abs{\phi_+} = \abs{\phi_-}$ after each stochastic dissipation step. To study the role of $\expval*{\hat{S}_z}$ conservation in more detail, we repeat the simulation of Fig. \ref{fig:Plot_Relaxation}(a), loosening the conservation law on $\expval*{\hat{S}_z}$ by omitting the above mentioned projection step. While $\expval*{\hat{S}_z}=0$ remains true on average, stochastic realizations may deviate from this strict conservation law. As shown in Fig. \ref{fig:conservation_law}, the correct steady state value $\overline{n}_p=1/4$ is recovered only when $\expval*{\hat{S}_z}$ is conserved at the level of individual trajectories. The breaking of $\expval*{\hat{S}_z}$ conservation within trajectories causes the pair number to converge towards the equilibrium value $\overline{n}_p=1/3$.
Hence, to correctly infer the dynamics of an isolated system using artificial dissipation, all symmetries of the Hamiltonian should be reflected within each individual trajectory.

\section{Dissipative dynamics}
\label{sec:dissipative}

\begin{figure}[tbp]
    \includegraphics[scale=1]{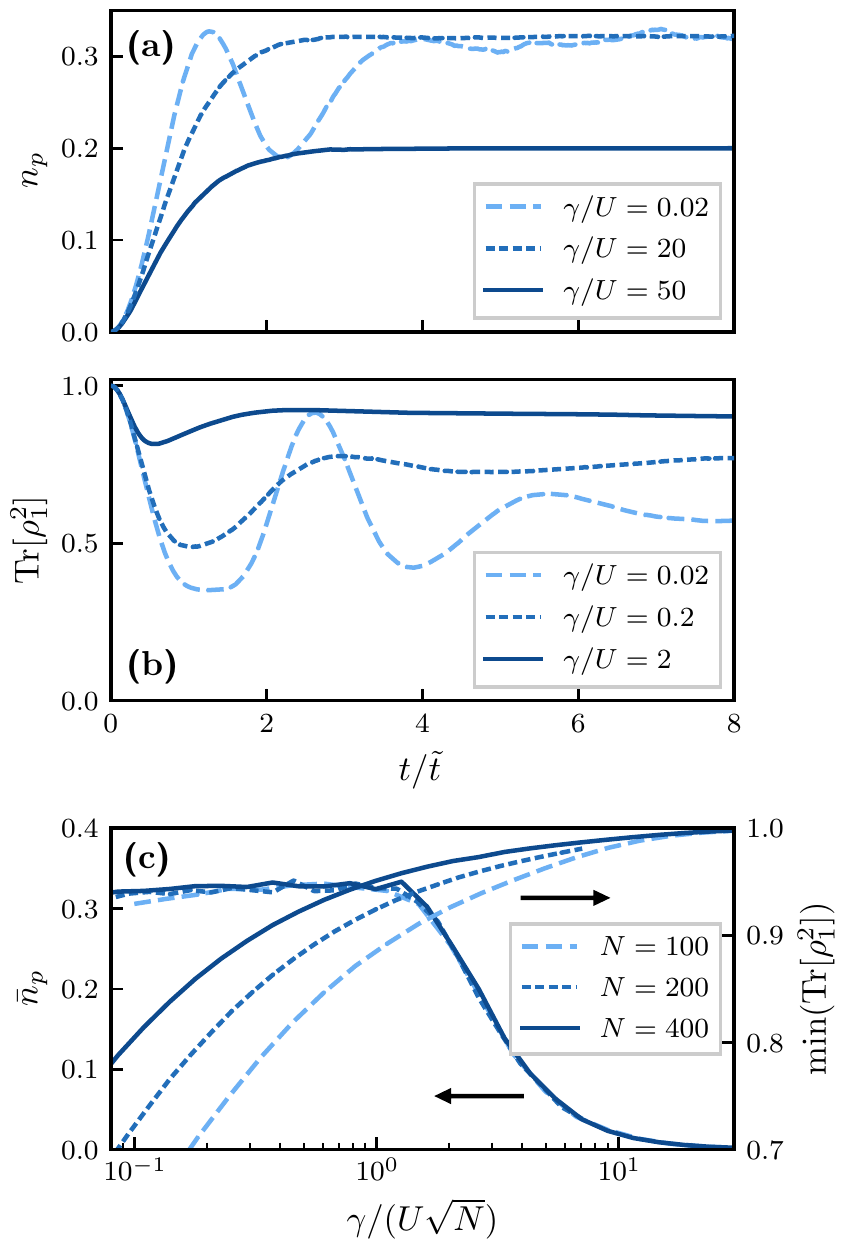}
    \caption{Time evolution of a dissipative spin-1 gas with initial $N=400$ quenched to $q/U = 0$, for different values of the dissipation rate $\gamma/U$. (a) Relaxation of the pair number due to spin mixing dynamics. (b) Average purity of individual trajectories. (c) Final pair fraction and minimal purity of an average trajectory as a function of $\gamma/(U\sqrt{N})$, indicating a window where a non-fragmented state may be observed.}
    \label{fig:Plot_Open_System}
\end{figure}

Having shown how the fragmented condensate can be represented as a mixture  of variational states, we may now ask whether such a single trajectory can be observed experimentally and how much the system should be disturbed to do so. 
To quantify this question, we simulate in Fig. \ref{fig:Plot_Open_System} the time evolution of an actual open system, dissipating into its environment after being quenched to spin degeneracy ($q/U=0$). This is achieved by including all dissipation terms in the time evolution of the Gaussian observables and omitting the projection steps which previously fixed the total particle number and angular momentum $\expval*{\hat{S}_z}$. As seen from Fig. \ref{fig:Plot_Open_System}(a), spin mixing dynamics eventually relax for any nonzero dissipation rate. At low $\gamma/U$, the internal dynamics are weakly affected by atom losses and the condensate evolves to a steady state characterized by $\bar{n}_p=1/3$. At high $\gamma/U$, rapid dissipation of atoms in the $\ket{0}$ state inhibits the interactions responsible for spin mixing, resulting in a lower $\bar{n}_p$.

Since the aim of adding particle losses is merely a defragmentation of the condensate, the dissipation strength should be weak enough to avoid significant depletion of the gas over experimentally relevant timescales (of the order $\tilde{t}$). On the other hand, the dissipation strength should be sufficiently strong to suppress Gaussian fluctuations in order to recover a trajectory with high condensate purity. Fig. \ref{fig:Plot_Open_System}(b) illustrates this consideration by showing the average purity of a single trajectory for several values of $\gamma/U$.
To quantify the compromise between particle losses and single trajectory purity, we plot in Fig. \ref{fig:Plot_Open_System}(c) the final fraction of pairs $\bar{n}_p$ together with the minimal purity during an average trajectory, for three different initial atom numbers.
The pair number curves for different atom numbers coincide after scaling the $x$-axis with $1/\sqrt{N}$ and show a sharp transition at $\gamma/U \approx \sqrt{N}$, where the dissipation rate of $\ket{0}$ atoms becomes comparable to the rate at which spin pairs are created. The minimal purity, on the other hand, increases smoothly with the dissipation rate and coincides for different $N$ after a rescaling of the $x$-axis with $\sqrt{N}\gamma/U$ (not shown).
Our results thus indicate that a window of high purity and weakly affected pair fraction exists within the range $1/\sqrt{N} \ll \gamma/U < \sqrt{N}$, which grows with increasing particle number. Note that our Gaussian trajectory calculations are not reliable in the regime of small $\gamma/U$ with large fluctuations.

Finally, it is worth pointing out that the evaluation of entanglement within a single trajectory of a dissipative system is possible only in a quantum trajectory approach which explicitly includes quantum fluctuations. By contrast, the TWA only provides access to the combined inter- and intra-trajectory variance of observables by inferring the density matrix through an average of realisations \cite{verstraelenGaussianQuantumTrajectories2018}.

\section{Conclusion and outlook}
\label{sec:conclusions}

We have shown how the range of applicability of a Gaussian variational Ansatz for the description of a quenched spinor gas in isolation can be extended to capture relaxation to a fragmented steady state.
In a quantum trajectory approach, the introduction of dissipation into a fictitious environment leads to the effective conversion of quantum fluctuations into classical mixedness of the many-body state, overcoming the incapacity of a single squeezed state to describe the highly entangled fragmented condensate. By simulating an actual open system, we have provided an indication for the strength of dissipation needed to observe a non-fragmented state in experiments.

Beyond the exactly solvable single-mode model presented here, the method of approximating nonequilibrium states as a classical mixture of Gaussian states provides a new outlook on the study of spatially resolved Bose gases, where recent experiments have uncovered rich physics in the thermalization process, including domain formation \cite{pruferObservationUniversalDynamics2018, jimenez-garciaSpontaneousFormationRelaxation2019} and the emergence of universal prethermal steady states \cite{eigenUniversalPrethermalDynamics2018, pruferObservationUniversalDynamics2018}.
While approximate solutions provided by TWA simulations are suitable to describe the dynamics at early and intermediate times \cite{barnettPrethermalizationQuenchedSpinor2011, schmiedBidirectionalUniversalDynamics2019}, they are incapable of capturing full thermalization because of the unavoidable ultraviolet catastrophe in classical field theories. In particular, the TWA fails to describe the dynamics of lowly occupied modes, such as those in the exponential tail of a thermal momentum distribution \cite{vanregemortelSpontaneousBeliaevLandauScattering2017}. By contrast, a Gaussian variational theory does not suffer from this limitation. The extension to kinetic thermalization in multimode systems is therefore a natural next challenge for the Gaussian trajectory approach and will be the subject of future work.

\begin{acknowledgments}
We thank Bertrand Evrard for kindly providing the experimental data from \cite{evrardManyBodyOscillationsThermalization2021,evrardCoherentSeedingDynamics2021} and additional information about the experiment.
L. F. gratefully acknowledges support from the Research Foundation - Flanders (FWO) through a Ph.D. fellowship fundamental research, project 11E8120N.
\end{acknowledgments}


%


\pagebreak
\onecolumngrid
\appendix
\numberwithin{figure}{section}

\section{Exact solution}
\label{appendix:hamiltonian}
\begin{figure}[htbp]
    \includegraphics[scale=1]{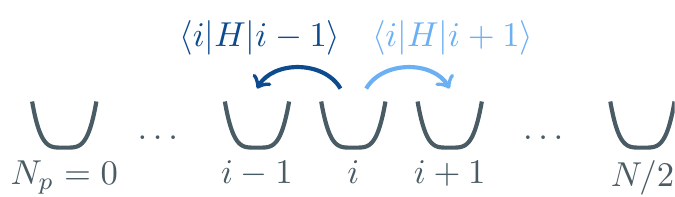}
    \caption{Single-particle hopping Hamiltonian to which the many-body system is mapped due to its conservation laws.}
    \label{fig:illustration_hopping}
\end{figure}

Conservation of particle number and angular momentum $\expval*{\hat{S}_z}$ restrict the evolution of the three-mode system to an $N/2+1$ dimensional subspace spanned by the basis of \emph{pair number states} $\ket{N_p}= \ket{N_p, N-2N_p, N_p}$, a subset of the Fock states $\ket{n_+,n_0,n_-}$, characterized by a well-defined number of particles occupying each mode. Within this subspace, the evolution of an arbitrary state $\ket{\psi}$ can be computed through the many-body Schr\"odinger equation
\begin{equation}
    i\partial_t\braket{n}{\psi} = \sum_{m=0}^{N/2} \matrixel{n}{\hat{\mathcal{H}}_S}{m} \braket{m}{\psi},
\end{equation}
where the matrix elements $ H_{nm} = \matrixel{n}{\hat{\mathcal{H}}_S}{m}$ are given by \cite{miasQuantumNoiseScaling2008}:
\begin{multline}
    H_{nm} = m\qty[2q + U  \qty(2(N-2m)-1)] \delta_{n,m} 
    + U m \sqrt{(N-2m+1)(N-2m+2)}\delta_{n,m-1} \\
    + U (m+1) \sqrt{(N-2m)(N-2m-1)} \delta_{n,m+1}.
\end{multline}
The tridiagonal shape of the Hamiltonian is a consequence of the processes pictured in Fig. \ref{fig:illustration_SMA}, through which one spin pair at a time is created or annihilated. The many-body system is therefore equivalent to a single particle on a chain with nearest-neighbour hopping amplitudes, as illustrated in Fig. \ref{fig:illustration_hopping}. 

Note that the basis of pair number states does not span the entire Hilbert space of a spin-1 gas with $\expval*{\hat{S}_z}=0$, since states such as $\ket{\psi} = (\ket{N,0,0} + \ket{0,0,N})/\sqrt{2}$ (in the Fock basis) cannot be represented as a superposition of pair number states. This poses no restriction on our discussion of the dynamics since all initial conditions correspond to \emph{coherent states} \cite{serafiniQuantumContinuousVariables2017, glauberCoherentIncoherentStates1963, weedbrookGaussianQuantumInformation2012}, expressible as a superposition of pair number states through
\begin{equation}
    \ket{\tilde{N}_p, \Delta}_{coh} = e^{-\frac{\tilde{N}_p}{2}}\sum_{n_p=0}^{N/2} \frac{\sqrt{\tilde{N}_p}^{n_p} e^{in_p \Delta}}{\sqrt{n_p!}} \ket{n_p}. \label{eq:coherent_state}
\end{equation}
Here, $\tilde{N}_p$ is the mean pair number and $\Delta= \theta_+ + \theta_- - 2\theta_0$ the relative phase of the coherent spin components, taken to be $\Delta = 0$ throughout the manuscript.
Since the matrix elements $H_{nm}$ connect pair number states only to other pair number states, the time evolution remains restricted to the subspace spanned by pair number states. 
The diagonal nature of the single-particle reduced density matrix $\rho^{(n,m)}_{1} = \expval*{\hat{a}_n^\dagger \hat{a}_m}/N$ is a direct consequence of this restriction. For any state $\ket{\psi}$ in the described subspace, $\hat{a}_n^\dagger \hat{a}_{m\neq n}\ket{\psi}$ is a superposition of strictly non pair number states, rendering $\matrixel{\psi}{\hat{a}_n^\dagger \hat{a}_{m\neq n}}{\psi}=0$ due to the orthogonality of Fock states. The reduced density matrix therefore simplifies to $\rho_{1} = \textnormal{diag}(n_p, 1-2n_p, n_p)$.
To obtain the correct ground state in Fig. \ref{fig:phasediagram}(a) of the main text, it too should be expressible as a superposition of pair number states. This was shown to be true at $q/U=0$ \cite{lawQuantumSpinsMixing1998}. In the high $q/U$ regime, our results agree with the Bogoliubov prediction detailed in Appendix \ref{appendix:bogoliubov} and a continuum approximation of the many-body Schr\"odinger equation \cite{sarloSpinFragmentationBose2013}.

\section{Bogoliubov expansion}
\label{appendix:bogoliubov}
The observed scaling of the pair number in the ground state and steady state at high $q/U$ can be derived by considering small spin excitations in a macroscopically occupied $\ket{0}$ condensate. The approximation $(\hat{a}_+, \hat{a}_0, \hat{a}_-) \approx (\hat{\delta}_+, \sqrt{N}, \hat{\delta_-})$ and expansion of the Hamiltonian up to quadratic order in fluctuation operators $\hat{\delta}_m$ lead to a Bogoliubov Hamiltonian, diagonalized by a transformation $\hat{\delta}_\pm = u\hat{b}_\pm + v^* \hat{b}_{\mp}^\dagger$ \cite{pitaevskiiBoseEinsteinCondensationSuperfluidity2016, evrardManyBodyOscillationsThermalization2021}.
Here, we introduced the coefficients 
\begin{equation}
u,v = \pm \sqrt{\frac{q+UN}{2\epsilon} \pm \frac{1}{2}},
\end{equation}
where $\epsilon = \sqrt{q(q+2UN)}$ is the dispersion of the non-interacting quasiparticles. The occupation of these Bogoliubov excitations is related to the number of spin pairs in the atomic basis through
\begin{equation}
    \expval*{\hat{\delta}_\pm^\dagger \hat{\delta}_\pm} = v^2 +  u^2 \expval*{\hat{b}_\pm^\dagger \hat{b}_\pm} + v^2 \expval*{\hat{b}_\mp^\dagger \hat{b}_\mp} 
    + uv(\expval*{\hat{b}_+ \hat{b}_-} + \expval*{\hat{b}_+^\dagger \hat{b}_-^\dagger}).
\end{equation}
The ground state corresponds to the Bogoliubov vacuum and thus contains $n_p = \abs{v}^2/N$ spin pairs, which scales as $(q/U)^{-1/2}$ in the regime $q/UN \ll 1 \ll q/U$. The same scaling was found in a continuum approximation of the many-body Schr\"odinger equation \cite{sarloSpinFragmentationBose2013}.
Conversely, the mean pair number following an instantaneous quench out of a $\ket{0}$ coherent state can be determined by considering that at $t=0$ no pairs have been created, $\expval*{\hat{\delta}_\pm^\dagger \hat{\delta}_\pm}= \expval*{\hat{\delta}_+ \hat{\delta}_-} =0$. The corresponding occupation of Bogoliubov excitations $\expval*{\hat{b}_\pm^\dagger \hat{b}_\pm} = v^2$ is a constant of motion, while the anomalous correlation $\expval*{\hat{b}_+ \hat{b}_-}$ trivially rotates under time evolution and cancels out when taking time averages \cite{vanregemortelPrethermalizationThermalizationCrossover2018}. The average pair number at late times is therefore given by $n_p = (u^2 + v^2 + 1)v^2/N$, scaling as $(q/U)^{-1}$ in the regime $q/UN \ll 1 \ll q/U$. 
Predictions of Bogoliubov theory become inaccurate in the ground state and dynamics at low $q/U$ \cite{evrardManyBodyOscillationsThermalization2021} or coherently seeded dynamics at high $q/U$ \cite{evrardCoherentSeedingDynamics2021}, owing to the no longer negligible occupation of the $\ket{\pm}$ modes.

\section{Gaussian trajectory theory}
\label{appendix:HFB}
\def\helta{\hat{\delta}}

Employing a Gaussian Ansatz for the field operators $\hat{a}_m \equiv \phi_m + \hat{\delta}_m$ ($m=-,0,+$), the many-body dynamics contained in the Hamiltonian \eqref{eq:spin_hamiltonian} are reduced to a closed system governing the evolution of the first moments $\phi_m = \expval*{\hat{a}_m}$ and all quadratic correlations of the fluctuations operators $\hat{\delta}_m$, satisfying $\expval*{\hat{\delta}_m}=0$. We denote these second central moments as

\begin{align}
    n_+ &= \expval*{\hat{\delta}_+^\dagger \hat{\delta}_+}, & c_+ &= \expval*{\hat{\delta}_+ \hat{\delta}_+}, & b_+ &= \expval*{\hat{\delta}_0 \hat{\delta}_+}, & d_+ &= \expval*{\hat{\delta}_0^\dagger \hat{\delta}_+}, \nonumber \\
    n_0 &= \expval*{\hat{\delta}_0^\dagger \hat{\delta}_0}, & c_0 &= \expval*{\hat{\delta}_0 \hat{\delta}_0}, & b_1 &= \expval*{\hat{\delta}_+ \hat{\delta}_-}, & d_1 &= \expval*{\hat{\delta}_+^\dagger \hat{\delta}_-}, \label{eq:appendix_moments} \\
    n_- &= \expval*{\hat{\delta}_-^\dagger \hat{\delta}_-}, & c_- &= \expval*{\hat{\delta}_- \hat{\delta}_-}, & b_- &= \expval*{\hat{\delta}_0 \hat{\delta}_-}, & d_- &= \expval*{\hat{\delta}_0^\dagger \hat{\delta}_-}. \nonumber
\end{align}

The Gaussian nature of fluctuations is reflected in the application of Wick's theorem to all higher order moments,
\begin{align}
    \expval*{\hat{a}\hat{b}\hat{c}\hat{d}} &= \expval*{\hat{a}\hat{b}}\expval*{\hat{c}\hat{d}} + \expval*{\hat{a}\hat{c}}\expval*{\hat{b}\hat{d}} + \expval*{\hat{a}\hat{d}}\expval*{\hat{b}\hat{c}}, \\
    \expval*{\hat{a}\hat{b}\hat{c}} &= \expval*{\hat{a}\hat{b}\hat{c}\hat{d}\hat{e}} = \dots = 0.
\end{align}
This Ansatz on the statistics closes the dynamics at the Gaussian level, restricting our description of the system to a small region of the full Hilbert space \cite{serafiniQuantumContinuousVariables2017, verstraelenGaussianQuantumTrajectories2018}.

To make the calculations tractable, the Hamiltonian is expanded in orders of fluctuations. Contributions of first and third order can be omitted as their effect vanishes under the Gaussian Ansatz, leading to
\begin{equation}
    \hat{\mathcal{H}}_S \approx E_0 + \hat{H}_2 + \hat{H}_4,
\end{equation}
where 
\begin{multline}
    E_0 = \frac{q}{\abs{U}}\qty(\abs{\phi_+}^2 + \abs{\phi_-}^2) + \frac{\sigma}{2} \Big( \abs{\phi_+}^4 + \abs{\phi_-}^4 - 2\abs{\phi_+}^2\abs{\phi_-}^2 \\ 
    + 2\abs{\phi_+}^2\abs{\phi_0}^2 + 2\abs{\phi_-}^2\abs{\phi_0}^2 + 2 \phi_+^* \phi_-^* \phi_0 \phi_0 + 2\phi_0^* \phi_0^* \phi_+ \phi_-  \Big),
\end{multline}

\begin{align}
    \hat{H}_2 =& \frac{q}{\abs{U}}\qty(\helta_+^\dagger \helta_+ + \helta_-^\dagger \helta_-) \nonumber\\
    &+ \frac{\sigma}{2}\Big( \phi_+^* \phi_+^* \helta_+ \helta_+ + \phi_+ \phi_+ \helta_+^\dagger \helta_+^\dagger + 4\abs{\phi_+}^2\helta_+^\dagger \helta_+  \Big)
    + \frac{\sigma}{2}\Big( \phi_-^* \phi_-^* \helta_- \helta_- + \phi_- \phi_- \helta_-^\dagger \helta_-^\dagger + 4\abs{\phi_-}^2\helta_-^\dagger \helta_-  \Big) \nonumber\\
    &-\sigma \Big(\abs{\phi_+}^2\helta_-^\dagger \helta_- + \abs{\phi_-}^2\helta_+^\dagger \helta_+ + \phi_+^* \phi_-^* \helta_+ \helta_- + \phi_+ \phi_- \helta_+^\dagger \helta_-^\dagger + \phi_+^* \phi_- \helta_-^\dagger \helta_+ + \phi_-^* \phi_+ \helta_+^\dagger \helta_- \Big) \nonumber\\
    &+ \sigma \Big(\abs{\phi_+}^2 \helta_0^\dagger \helta_0 + \abs{\phi_0}^2 \helta_+^\dagger \helta_+ 
    + \phi_+^* \phi_0^* \helta_+ \helta_0 + \helta_+^\dagger \helta_0^\dagger \phi_+ \phi_0 + \phi_+^* \phi_0 \helta_0^\dagger \helta_+ + \phi_0^* \phi_+ \helta_+^\dagger \helta_0 \Big) \nonumber\\
    &+\sigma\Big( \abs{\phi_-}^2 \helta_0^\dagger \helta_0 + \abs{\phi_0}^2 \helta_-^\dagger \helta_- + \phi_-^* \phi_0^* \helta_- \helta_0 + \helta_-^\dagger \helta_0^\dagger \phi_- \phi_0 + \phi_-^* \phi_0 \helta_0^\dagger \helta_- + \phi_0^* \phi_- \helta_-^\dagger \helta_0 \Big) \nonumber\\
    &+ \sigma \Big(\phi_0^* \phi_0^* \helta_+ \helta_- + \helta_0^\dagger \helta_0^\dagger \phi_+ \phi_- + 2\phi_0^*\phi_+ \helta_0^\dagger \helta_- +  2\phi_0^*\phi_- \helta_0^\dagger \helta_+ \Big) \nonumber\\
    &+ \sigma \Big( \phi_+^* \phi_-^* \helta_0 \helta_0 + \phi_0 \phi_0 \helta_+^\dagger \helta_-^\dagger + 2 \phi_+^* \phi_0 \helta_-^\dagger \helta_0 + 2 \phi_-^* \phi_0 \helta_+^\dagger \helta_0 \Big),
\end{align}

\begin{equation}
    \hat{H}_4 = \frac{\sigma}{2} \Big( \helta_+^\dagger \helta_+^\dagger \helta_+ \helta_+ + \helta_-^\dagger \helta_-^\dagger \helta_- \helta_- - 2 \helta_+^\dagger \helta_-^\dagger \helta_+ \helta_- + 2 \helta_+^\dagger \helta_0^\dagger \helta_+ \helta_0 + 2 \helta_-^\dagger \helta_0^\dagger \helta_- \helta_0 + 2 \helta_0^\dagger \helta_0^\dagger \helta_+ \helta_- + 2 \helta_+^\dagger \helta_-^\dagger \helta_0 \helta_0 \Big).
\end{equation}

The evolution of an observable under heterodyne unraveling \eqref{eq:heterodyne_lindblad} can be rewritten as
\begin{equation}
    \dd \expval*{\hat{O}} = \, - i \expval{\commutator{\hat{O}}{\hat{\mathcal{H}}_S}} \dd t + \mathcal{D}(\expval*{\hat{O}}), \label{eq:appendix_heterodyne}
\end{equation}
where the dissipator $\mathcal{D}$ contains all effects of dissipation. 
As the first moments $\phi_m$ of the operators $\hat{a}_m$ were explicitly substituted in the expanded Hamiltonian, the above commutator can no longer be used to find their time evolution. Instead, the time evolution of these classical wave functions is equivalently found by treating $\phi_m$ and $\phi_m^*$ as canonical variables of a classical Hamiltonian \cite{colussiCumulantTheoryUnitary2020}, resulting in
\begin{equation}
    \dd \phi_m = \, - i \expval{\pdv{(E_0 + \hat{H}_2)}{\phi_m^*}} \dd t + \mathcal{D}(\phi_m). \label{eq:appendix_heterodyne_phi}
\end{equation}
For the second moments, Eq. \eqref{eq:appendix_heterodyne} reduces to
\begin{equation}
    \dd \expval*{\hat{\delta}^{(\dagger)}_m\hat{\delta}_n} = \, - i \expval{\commutator{\hat{\delta}^{(\dagger)}_m\hat{\delta}_n}{(\hat{H}_2 + \hat{H}_4)}} \dd t + \mathcal{D}(\expval*{\hat{\delta}^{(\dagger)}_m\hat{\delta}_n}), \label{eq:appendix_heterodyne_fluct}
\end{equation}

\subsection{Hamiltonian evolution} 
For the mean field modes $\phi_m=\expval*{\hat{a}_m}$, one finds the Hamiltonian part of \eqref{eq:appendix_heterodyne_phi} as
\begin{multline}
   \expval{\pdv{(E_0 + \hat{H}_2)}{\phi_+^*}} = q\phi_+ + U\qty[(\abs{\phi_0}^2 + \abs{\phi_+}^2 -\abs{\phi_-}^2)\phi_+ + \phi_-^*\phi_0\phi_0] \\
    + U \qty[(n_0 + 2n_+ -n_-)\phi_+ + \phi_0^* b_+ + \phi_0 d_+ + \phi_-^* c_0 + 2 \phi_0 d_-^* + \phi_+^* c_+ - \phi_-^* b_1 - \phi_- d_1^*],
\end{multline}
\begin{multline}
    \expval{\pdv{(E_0 + \hat{H}_2)}{\phi_0^*}} = U \qty[(\abs{\phi_+}^2 + \abs{\phi_-}^2)\phi_0 + 2\phi_0^*\phi_+ \phi_-] \\
    + U\qty[(n_+ + n_-)\phi_0 + \phi_+^* b_+ + \phi_-^* b_- + \phi_+ d_+^* + \phi_- d_-^* +2\phi_0^* b_1 + 2\phi_+ d_- + 2\phi_- d_+],
\end{multline}
\begin{multline}
    \expval{\pdv{(E_0 + \hat{H}_2)}{\phi_-^*}} = q\phi_- + U \qty[(\abs{\phi_0}^2 + \abs{\phi_-}^2 -\abs{\phi_+}^2)\phi_- + \phi_+^*\phi_0\phi_0] \\
    + U \qty[(n_0 + 2n_- -n_+)\phi_- + \phi_0^* b_- + \phi_0 d_- + \phi_+^* c_0 + 2 \phi_0 d_+^* + \phi_-^* c_- - \phi_+^* b_1 - \phi_+ d_1],
\end{multline}
where the first and second lines denote the contributions of $E_0$ and $\hat{H}_2$, respectively.

The evolution of the quadratic moments \eqref{eq:appendix_moments} under $\hat{H}_2$ is given by:
\begin{multline}
   \expval{\commutator{\hat{n}_+}{\hat{H}_2}}  = 2U i \Im \qty{(\phi_0^*\phi_+ + 2 \phi_-^* \phi_0) d_+^* - \phi_-^* \phi_+ d_1 + \phi_+ \phi_0 b_+^* + \phi_+ \phi_+ c_+^* + (\phi_0 \phi_0 - \phi_+ \phi_-)b_1^*}, \hfill
\end{multline}
\begin{multline}
   \expval{\commutator{\hat{n}_0}{\hat{H}_2}}  = 2U i \Im \qty{(\phi_0 \phi_+^* + 2 \phi_0^* \phi_-)d_+ + (\phi_0 \phi_-^* + 2 \phi_0^* \phi_+)d_- + 2\phi_+ \phi_- c_0^* + \phi_0 \phi_+ b_+^* + \phi_0 \phi_- b_-^*}, \hfill
\end{multline}
\begin{multline}
   \expval{\commutator{\hat{n}_-}{\hat{H}_2}}  = 2U i \Im \qty{(\phi_0^*\phi_- + 2 \phi_+^* \phi_0) d_-^* - \phi_+^* \phi_- d_1^* + \phi_- \phi_0 b_-^* + \phi_- \phi_- c_-^* + (\phi_0 \phi_0 - \phi_- \phi_+)b_1^*}, \hfill
\end{multline}
\begin{multline}
   \expval{\commutator{\hat{c}_+}{\hat{H}_2}}  = 2 \qty[q + U(\abs{\phi_0}^2 + 2 \abs{\phi_+}^2 - \abs{\phi_-}^2)]c_+ \\
   + 2U\qty[(\phi_0^* \phi_+ + 2 \phi_-^* \phi_0)b_+ - \phi_-^* \phi_+ b_1 + \phi_+ \phi_0 d_+ + \phi_+ \phi_+ (n_+ + 1/2) + (\phi_0 \phi_0 - \phi_+ \phi_-)d_1^*]
\end{multline}
\begin{multline}
   \expval{\commutator{\hat{c}_0}{\hat{H}_2}}  = 2U \qty[\abs{\phi_+}^2 + \abs{\phi_-}^2]c_0 \\
   + 2U\qty[(\phi_0 \phi_+^* + \phi_0^* \phi_-)b_+ + (\phi_0 \phi_-^* + 2\phi_0^* \phi_+)b_- + 2\phi_+ \phi_-(2n_0+1)  +  \phi_0 \phi_+ d_+^* + \phi_0 \phi_- d_-^* ] 
\end{multline}
\begin{multline}
   \expval{\commutator{\hat{c}_-}{\hat{H}_2}} = 2 \qty[q + U(\abs{\phi_0}^2 + 2 \abs{\phi_-}^2 - \abs{\phi_+}^2)]c_- \\
   + 2U\qty[(\phi_0^* \phi_- + 2 \phi_+^* \phi_0)b_- - \phi_+^* \phi_- b_1 + \phi_- \phi_0 d_- + \phi_- \phi_- (n_- + 1/2) + (\phi_0 \phi_0 - \phi_+ \phi_-)d_1]
\end{multline}
\begin{multline}
   \expval{\commutator{\hat{b}_+}{\hat{H}_2}}  =  \qty[q + U(\abs{\phi_0}^2 + 3 \abs{\phi_+}^2) ]b_+ \\
   + U\Big[(\phi_0 \phi_+^* + 2\phi_0^* \phi_-) c_+ + (\phi_0 \phi_-^* + 2 \phi_0^* \phi_+)b_1 + 2\phi_+\phi_-d_+  + \phi_0 \phi_- d_1^* + (\phi_0^* \phi_+ + 2\phi_-^* \phi_0)c_0 \\
    - \phi_-^* \phi_+ b_- + \phi_+ \phi_0(n_0 + n_+ +1) + \phi_+ \phi_+ d_+^* + (\phi_0 \phi_0 - \phi_+ \phi_-) d_-^* \Big]
\end{multline}
\begin{multline}
   \expval{\commutator{\hat{b}_1}{\hat{H}_2}}  = \qty[2q + U(2\abs{\phi_0}^2 +  \abs{\phi_+}^2 + \abs{\phi_-}^2) ]b_1 \\
   + U\Big[(\phi_0^* \phi_+ + 2\phi_-^* \phi_0) b_- + (\phi_0^* \phi_- + 2 \phi_+^* \phi_0)b_+ - \phi_+^*\phi_- c_+ - \phi_-^*\phi_+ c_- + \phi_0\phi_+ d_- + \phi_0\phi_- d_+ \\
   + \phi_+ \phi_+ d_1 + \phi_- \phi_- d_1^* + (\phi_0\phi_0 - \phi_+ \phi_-)(n_+ + n_- + 1) \Big]
\end{multline}
\begin{multline}
   \expval{\commutator{\hat{b}_-}{\hat{H}_2}} = \qty[q + U(\abs{\phi_0}^2 + 3 \abs{\phi_-}^2) ]b_- \\
   + U\Big[(\phi_0 \phi_-^* + 2\phi_0^* \phi_+) c_- + (\phi_0 \phi_+^* + 2 \phi_0^* \phi_-)b_1 + 2\phi_+\phi_-d_-  + \phi_0 \phi_+ d_1 + (\phi_0^* \phi_- + 2\phi_+^* \phi_0)c_0 \\
    - \phi_+^* \phi_- b_+ + \phi_- \phi_0(n_0 + n_- +1) + \phi_- \phi_- d_-^* + (\phi_0 \phi_0 - \phi_+ \phi_-) d_+^* \Big]
\end{multline}
\begin{multline}
   \expval{\commutator{\hat{d}_+}{\hat{H}_2}}  =  \qty[q + U(\abs{\phi_0}^2 + \abs{\phi_+}^2 - 2\abs{\phi_-}^2) ]d_+ \\
   + U\Big[(\phi_0^* \phi_+ + 2\phi_-^* \phi_0) n_0 - \phi_-^* \phi_+ d_- + \phi_+ \phi_0 c_0^* + \phi_+ \phi_+ b_+^* + (\phi_0 \phi_0 - \phi_+ \phi_-) b_-^* \Big]\\
   - U\Big[ (\phi_0^*\phi_+ + 2\phi_0\phi_-^*)n_+ + (\phi_0^* \phi_- + 2\phi_0 \phi_+^*)d_1^* + 2\phi_+^*\phi_-^* b_+ + \phi_0^* \phi_+^* c_+ + \phi_0^* \phi_-^* b_1 \Big]
\end{multline}
\begin{multline}
   \expval{\commutator{\hat{d}_1}{\hat{H}_2}}  = U\Big[3(\abs{\phi_-}^2 -\abs{\phi_+}^2) d_1 + (\phi_0^* \phi_- + 2\phi_+^* \phi_0)d_+^* + \phi_+^* \phi_-(n_- - n_+) + \phi_- \phi_0 b_+^* + \phi_- \phi_- b_1^*\\
   + (\phi_0 \phi_0 - \phi_+ \phi_-)c_+^* - (\phi_0\phi_+^* + 2\phi_-\phi_0^*)d_- - \phi_+^* \phi_0^* b_- - \phi_+^* \phi_+^* b_1 - (\phi_0^* \phi_0^* - \phi_+^* \phi_-^*)c_- \Big]
\end{multline}
\begin{multline}
   \expval{\commutator{\hat{d}_-}{\hat{H}_2}}  =  \qty[q + U(\abs{\phi_0}^2 + \abs{\phi_-}^2 - 2\abs{\phi_+}^2) ]d_- \\
   + U\Big[(\phi_0^* \phi_- + 2\phi_+^* \phi_0) n_0 - \phi_+^* \phi_- d_+ + \phi_- \phi_0 c_0^* + \phi_- \phi_- b_-^* + (\phi_0 \phi_0 - \phi_+ \phi_-) b_+^* \Big]\\
   - U\Big[ (\phi_0^*\phi_- + 2\phi_0\phi_+^*)n_- + (\phi_0^* \phi_+ + 2\phi_0 \phi_-^*)d_1 + 2\phi_+^*\phi_-^* b_- + \phi_0^* \phi_-^* c_- + \phi_0^* \phi_+^* b_1 \Big]
\end{multline}

The Wick-contracted correction to their time evolution coming from $\hat{H}_4$ is given by:
\begin{align}
   \expval{\commutator{\hat{n}_+}{\hat{H}_4}}  &= 2U i \Im \qty{b_1^* c_0 + 2 d_+^* d_-^*}, \\
   \expval{\commutator{\hat{n}_0}{\hat{H}_4}}  &= 2U i \Im \qty{2c_0^* b_1 + 4d_+ d_-}, \\
   \expval{\commutator{\hat{n}_-}{\hat{H}_4}}  &= 2U i \Im \qty{b_1^* c_0 + 2 d_-^* d_+^*}, \\
   \expval{\commutator{\hat{c}_+}{\hat{H}_4}}  &= Uc_+ + 2U \Big[3 n_+ c_+ - 2 d_1^* b_1 - n_- c_+ + n_0 c_+ + 2d_+ b_+ + d_1^* c_0 + 2d_-^*b_+ \Big], \\
   \expval{\commutator{\hat{c}_0}{\hat{H}_4}}  &= 2U \Big[ n_+ c_0 + 2 d_+^*b_+ + n_-c_0 + 2d_-^*b_- + 2n_0b_1 + 2d_+ b_- + 2d_-b_+ + b_1 \Big], \\
   \expval{\commutator{\hat{c}_-}{\hat{H}_4}}  &= Uc_- + 2U \Big[3 n_- c_- - 2 d_1 b_1 - n_+ c_- + n_0 c_- + 2d_- b_- + d_1 c_0 + 2d_+^*b_- \Big], \\
   \expval{\commutator{\hat{b}_+}{\hat{H}_4}}  &= U \Big[4n_+ b_+ + 2d_+^*c_+ + 4d_+b_1 + 2d_-c_+ + 2n_0b_+ + d_+ c_0 + 3d_-^*c_0 + b_+ \Big], \\
   \expval{\commutator{\hat{b}_1}{\hat{H}_4}}  &= U \Big[-b_1 + c_0 + 2n_0b_1 + 2d_+ b_- + 2d_-b_+ + n_-c_0 + 2d_-^*b_- + n_+c_0 + 2d_+^*b_+ \Big], \\
   \expval{\commutator{\hat{b}_-}{\hat{H}_4}}  &= U \Big[4n_- b_- + 2d_-^*c_- + 4d_-b_1 + 2d_+c_- + 2n_0b_- + d_- c_0 + 3d_+^*c_0 + b_- \Big], \\
   \expval{\commutator{\hat{d}_+}{\hat{H}_4}}  &= U \Big[2n_0d_+ + c_0^*b_+ + 2n_0d_-^* + b_-^*c_0 - 2b_-^*b_1 - 2d_+ n_- - 2d_-d_1^* - 2 b_1^*b_+ - 2d_+^*d_1^* - 2n_+d_-^*], \\
   \expval{\commutator{\hat{d}_1}{\hat{H}_4}}  &= U \Big[2b_1^*c_- + 4n_-d_1 - 2c_+^*b_1 - 4n_+d_1 + c_+^*c_0 + 2d_+^*d_+^* - c_0^*c_- - 2d_- d_- \Big], \\
   \expval{\commutator{\hat{d}_-}{\hat{H}_4}}  &= U \Big[2n_0d_- + c_0^*b_- + 2n_0d_+^* + b_+^*c_0 - 2b_+^*b_1 - 2d_- n_+ - 2d_+d_1 - 2 b_1^*b_- - 2d_-^*d_1 - 2n_-d_+^*].
\end{align}
Being of fourth order in fluctuation operators, this contribution of $\hat{H}_4$ to the dynamics is negligible in the trajectory description, as fluctuations are suppressed by construction. However, their effect is significant in the unaltered HFB theory, where fluctuations eventually dominate the dynamics.

\subsection{Dissipation}
Referring to Eq. \eqref{eq:heterodyne_lindblad} in the main text, the dissipator is given by
\begin{equation}
    \mathcal{D}(\expval*{\hat{O}}) = 
    - \frac{\gamma}{2}\sum_m \qty( \expval{\anticommutator{\hat{a}_m^\dagger \hat{a}_m}{\hat{O}}} - 2 \expval{\hat{a}_m^\dagger \hat{O} \hat{a}_m} )\dd t
    + \sqrt{\gamma} \sum_m \qty( \expval{\hat{a}_m^\dagger(\hat{O}- \expval*{\hat{O}})}\dd Z_m+ c.c. ).
    \label{eq:appendix_dissipator}
\end{equation}

For the first moments $\phi_m=\expval*{\hat{a}_m}$, one finds
\begin{align}
    \mathcal{D}(\phi_+) &= -\frac{\gamma}{2}\phi_+ \dd t + \sqrt{\gamma}\qty(n_+ \dd Z_+ + c_+ \dd Z_+^* + d_+ \dd Z_0 + b_+ \dd Z_0^* + d_1^* \dd Z_- + b_1 \dd Z_-^*), \\
    \mathcal{D}(\phi_0) &= -\frac{\gamma}{2}\phi_0 \dd t + \sqrt{\gamma}\qty(d_+^* \dd Z_+ + b_+ \dd Z_+^* + n_0 \dd Z_0 + c_0 \dd Z_0^* + d_-^* \dd Z_- + b_- \dd Z_-^*), \\
    \mathcal{D}(\phi_-) &= -\frac{\gamma}{2}\phi_- \dd t + \sqrt{\gamma}\qty(d_1 \dd Z_+ + b_1 \dd Z_+^* + d_- \dd Z_0 + b_- \dd Z_0^* + n_- \dd Z_- + c_- \dd Z_-^*).
\end{align}

The dissipator of the second moments $\expval*{\hat{\delta}_m^{(\dagger)}\hat{\delta}_n} = \expval*{\hat{a}_m^{(\dagger)}\hat{a}_n} - \phi_m^{(*)}\phi_n$ is found by first calculating $\mathcal{D}(\phi_m^{(*)}\phi_n)$ using It\={o}'s lemma ($\dd \phi_m^{(*)}\phi_n = \phi_m^{(*)}\dd \phi_n + \phi_n \dd \phi_m^{(*)} + \dd \phi_m^{(*)} \dd \phi_n$), and subtracting it from $\mathcal{D}(\expval*{\hat{a}_m^{(\dagger)}\hat{a}_n})$. The stochastic part is found to vanish, leaving only a deterministic dissipation:

\begin{align}
    \mathcal{D}(n_+) &= -\gamma \qty[n_+(n_+ + 1) + \abs{c_+}^2 + \abs{d_+}^2 + \abs{b_+}^2 + \abs{d_1}^2 + \abs{b_1}^2 ]\dd t, \\
    \mathcal{D}(n_0) &= -\gamma \qty[n_0(n_0 + 1) + \abs{c_0}^2 + \abs{d_+}^2 + \abs{b_+}^2 + \abs{d_-}^2 + \abs{b_-}^2 ]\dd t, \\
    \mathcal{D}(n_-) &= -\gamma \qty[n_-(n_- + 1) + \abs{c_-}^2 + \abs{d_-}^2 + \abs{b_-}^2 + \abs{d_1}^2 + \abs{b_1}^2 ]\dd t, \\
    \mathcal{D}(c_+) &= -\gamma \qty[c_+(2n_+ + 1) + 2d_+b_+ + 2b_1 d_1^*]\dd t, \\
    \mathcal{D}(c_0) &= -\gamma \qty[c_0(2n_0 + 1) + 2d_+^*b_+ + 2d_-^*b_-]\dd t, \\
    \mathcal{D}(c_-) &= -\gamma \qty[c_-(2n_- + 1) + 2d_-b_- + 2b_1 d_1]\dd t, \\
    \mathcal{D}(b_+) &= -\gamma \qty[b_+(n_+ + n_0 + 1) + d_+^*c_+  + c_0 d_+ + d_-^*b_1 + b_-d_1^*]\dd t, \\
    \mathcal{D}(b_1) &= -\gamma \qty[b_1(n_+ + n_- + 1) + c_+d_1 + c_-d_1^* + d_+b_- + b_+ d_-]\dd t, \\
    \mathcal{D}(b_-) &= -\gamma \qty[b_-(n_- + n_0 + 1) + d_-^*c_-  + c_0 d_- + d_+^*b_1 + b_+d_1]\dd t, \\
    \mathcal{D}(d_+) &= -\gamma \qty[d_+(n_+ + n_0 + 1) + b_+^* c_+ + c_0^*b_+ + d_-d_1^* + b_-^* b_1]\dd t, \\
    \mathcal{D}(d_1) &= -\gamma \qty[d_1(n_+ + n_- + 1) + c_+^*b_1 + c_-b_1^* + d_+^*d_- + b_+^*b_-]\dd t, \\
    \mathcal{D}(d_-) &= -\gamma \qty[d_-(n_- + n_0 + 1) + b_-^* c_- + c_0^*b_- + d_+d_1 + b_+^* b_1]\dd t.
\end{align}

\end{document}